# A Quantum-like Model for Predicting Human Decisions in the Entangled Social Systems

Aghdas Meghdadi, M. R. Akbarzadeh-T., and Kurosh Javidan

*Abstract*—Human-centered systems of systems such as social networks, the Internet of Things, or healthcare systems are growingly becoming significant facets of modern life. Realistic models of human behavior in such systems play an essential role in their accurate modeling and prediction. Nevertheless, human behavior under uncertainty often violates the predictions by the conventional probabilistic models. Recently, quantum-like decision theories have shown a considerable potential to explain the contradictions in human behavior by applying quantum probabilities. But providing a quantum-like decision theory that could *predict* rather than *describe* the current state of human behavior is still one of the unsolved challenges. The fundamental contribution of this work is introducing the concept of entanglement from quantum information theory to Bayesian networks. This concept leads to an *entangled quantum-like Bayesian network*, in which each human is a part of the entire society. Accordingly, society's effect on the dynamic evolution of the decision-making process, which is less often considered in decision theories, is modeled by entanglement measures. To reach this aim, we introduce a quantum-like witness and find the relationship between this witness and the famous concurrence entanglement measure. The proposed predictive entangled quantum-like Bayesian network (PEQBN) is evaluated on 22 experimental tasks. Results confirm that PEQBN provides more realistic predictions of human decisions under uncertainty when compared with classical Bayesian networks and three recent quantum-like approaches.

*Index Terms*— Entanglement, Quantum Physics, Human behavior, Social systems, Quantum-like decision-making, Bayesian networks.

## I. Introduction

Modeling complex human-centered system of systems such as smart cities, the Internet of Things, social networks, healthcare, or urban traffic is one of the most challenging problems in modern life. In the control process of such human-inclusive multi-agent systems, which deal with large amounts of uncertain data, models of human selection behavior serve a primary role. Although various approaches have been proposed for modeling human decision-making, most of them focus only on maximizing personal profit, rational and competitive behaviors. While other elements, such as collective benefit and cooperation among agents, stress, and other psychological pressures, that also affect the human decision-making process are largely neglected.

Human decision-making may be approached from different points of view, such as psychology [1], medicine [2], economics [3][4], engineering [5], and intelligent control [6][7]. Some modeling frameworks in this domain are Bayesian networks (BN) [8][9], expected utility (EU) [10][11], Markov decision theory [12][13], game theory [14][15], Dempster-Shafer theory [16][17] and fuzzy decision-making [18][19]. While these models and their extended variants have achieved considerable success in handling uncertainty [17][20], several important paradoxes such as violation of total probability law (TPL) [21] and the sure-things principle (STP) [22] on human behavior evade a solution [23][24]. These frameworks also cannot explain order effects [25], conjunction, and disjunction error [24] and fail to explain the well-known Allais and Ellsberg paradoxes [26], as well as new studies in social society [27].

The similarity between these paradoxes and some contradictions in physics has recently led to a series of quantum-like decision theories [28][23]. It is noted that quantum-like decision theories, including our proposed model, are not about modeling the decision-making process in the brain nor about using quantum computers. These models do not deal with actual quantum systems. Instead, the mathematical structure of quantum probability (QP) in the Hilbert space, as a generalization of classical probability (CP), is applied to the modeling of human selection behavior [29]. Recently QP achieved many successes in explaining the violation of CP in areas outside of physics [29], which is referred to as the "Second Quantum Revolution" [27][30].

Applying QP in decision-making theory is first introduced by Nakagomi [31]. He considered structural decision-making similar to the quantum measurement process, which can change the system's state. This idea was developed by other researchers in decision-making [32][29], cognition [23][33], Judgment [34][35], and reasoning [36][37], which have been comprehensively reviewed in [38]. Busemeyer et al. [32] in 2006 presented a quantum dynamical model similar to the Markov decision theory. They describe the evolution of complex-valued probability amplitude over time, similar to the Schrödinger equation. Then, Yukalov et al. [29] presented quantum decision theory (QDT). This model considered the objective probability similar to expected utility, as well as subjective interference. They defined a quantum parameter and set the static interference term equal to 0.25 to describe some of the mentioned paradoxes. Recently, Xiao [39] proposed the quantum-inspired complex mass function for Dempster-Shafer

Aghdas Meghdadi and M. R. Akbarzadeh-T are with the Department of Electrical Engineering, Center of Excellence on Soft Computing and Intelligent Information Processing, Ferdowsi University of Mashhad, Mashhad, Iran. (e-mail: (a.meghdadi@mail.um.ac.ir, e-mail: akbazar@um.ac.ir).

Kourosh Javidan is with the Department of Physics, Ferdowsi University of Mashhad, Mashhad, Iran. (e-mail: javidan@um.ac.ir).

theory to predict interference effects on human decision-making. One of the successful models in this domain is Quantum-like BN developed by Moreira et al. [28]. Classical BNs are the most applied structure to predict decisions by considering interconnected nodes and cause-and-effect relationships. However, the predictions of BNs in multi-agent social systems under uncertainty do not correspond well to reality [40]. In quantum-like BNs, the evolution of different decisions on a particular issue is simulated by wave functions overlapping with each other.

Here, we focus on presenting a predictive model of human selection behavior, i.e., the future likeliness of a decision, between 2 choices under uncertainty. We simulated a binary decision-making experiment by presenting some new concepts in a quantum-like BN structure. The idea of using QP in the BNs structure is introduced by Tucci in 1995 to solve a physical problem [41]. He stated that any classical BN could be expressed as an infinite number of quantum BNs. In such a structure, the probability value ($Pr_n$) in the BN is replaced by a suitable complex wave function as $\sqrt{Pr_n} e^{i\theta_n}$ where $\theta_n$ is a phase factor. Suppose the system's wave function contains over one eigenfunction of interacting parts. In that case, we may find destructive or constructive interferences terms due to phase parameters in the wave functions. One can explain most of the mentioned paradoxes in this section by considering proper values for interference terms. However, this approach does not provide any method (or logical conjecture) for determining (or at least estimating) phase factors and interference terms. Estimating the interference value is the fundamental challenge in quantum-like decision theories, leading us to present a predictive model.

Moreira et al. [28] presented one of the rare predictive quantum-like BN (QBN) models. They introduced a heuristic function to estimate the interference term. Introducing such heuristic procedures may be meaningful when large amounts of data have been processed. Because of the limited number of data in this domain, defining such a function with no justifications is a drawback of this method. Also, the interference term in this method can be selected from piecewise heuristic function. Besides, despite presenting a quantum-like model in [28], the classical view is used to define basis vectors and solve a quantum problem. After that, Huang et al. [42] tried to present a logical justification for this challenge by applying physical Deng entropy [43] instead of a heuristic function. The term entropy was initially introduced for measuring the uncertainty in thermodynamics and then extended to information theory by Shannon [44]. This concept is now used in various fields such as decision making domain [45][46]. But the method proposed by Huang et al. [42] has some drawbacks because of the improper use of physical function, Deng entropy, and assigning values greater than 1 to the cosine function in some situations. This drawback is discussed in section IV.D with more details. So, although quantum-like BNs have achieved considerable success in justifying paradoxes by considering suitable value for interference terms, presenting a predictive quantum-like BN is still one of the unsolved challenges.

This paper tries to maintain the advantages of previously explained methods and reduce their disadvantages by taking inspiration from the quantum information theory (QIT). The main novelty of our approach is introducing entangled quantum-like BNs based on QIT concepts. A composite system in QIT is entangled when unknown relations link the individual components as a single entity. We consider entangled nodes in the basic structure of quantum-like BN. Hence, in addition to direct relations modeled by arcs in BN, some unknown relations, such as the cooperation mechanism, are modeled.

Motivated by this issue, human society is considered a multi-agent system in which each agent is assumed as a part of the entire system, and cooperation between agents is considered as well as competition behaviors. In fact, modeling cooperative behavior in human societies has become a major scientific challenge [47][48]. Recently, statistical physics and network techniques are known as successful tools for studying cooperation in social games [49][50]. Perc et al. [47] reviewed the main methods of statistical physics for modeling interactions among individuals through collective behavior.

On the other hand, cognitive and neurophysiological outcomes confirm that human decisions are made in two ways: i) interpreting the information for each option separately, or by ii) thinking about all options together and comparing the differences between collected information [51]. Inspired by this theory and QIT, the dynamical evolution for selecting choices is simulated as the trajectory of electrons as either particles or waves under certain or uncertain situations, respectively. If there is no uncertainty, the dynamical evolution of selecting each option interprets separately, which agrees with the classical theories. Alternatively, in the presence of uncertainty, when a decision-maker (DM) has not reached a final decision, we consider the superposition of different choices by modeling overlapping wave functions. Due to the existence of phase parameters in the wave functions, we may find different possibilities for destructive or constructive interferences. By considering interference effects, most of the paradoxes in the decision-making domain can be explained.

Another innovation of our study is relating the phase parameters to environmental conditions and estimating the effect of society on the decision-making process by entanglement measures. So interference term in our method is estimated by two well-known measures and a novel witness of quantum entanglement instead of considering only a heuristic function in other methods [28] or static value presented in [29]. Specifically, a new quantum-like witness is defined besides applying concurrence and Shannon entropy measures in our model.

Hence, we propose a predictive entangled quantum-like Bayesian network (PEQBN) inspired by neurological evidence and quantum concepts. The superiority of PEQBN is shown by evaluating this model on the prisoner's dilemma (PD) and two-stage gambling games as well as a recent document-relevant judgment task in which participants must judge about relating documents to specific queries [52]. The rest of this paper is organized as follows. Section II presents some preliminaries, including an example of a violation of CP and introducing QP. Section III presents the proposed predictive entangled quantum-like BN (PEQBN) approach. In Section IV, the proposed model is evaluated on 22 experiments of 3 decision-making tasks and compared to BN and three recently published quantum-like decision models [28][29][42]. Finally, Section V offers conclusions and suggestions for future works.

## II. Preliminaries

### A. Paradoxes in Classical Decision-Theory

There are several well-known paradoxes in the decision-making domain that violate the principles of CP. The common feature of these paradoxes is the existence of different kinds of uncertainty. The "Linda problem" and "Gallup Poll question" are examples of conjunction and disjunction error and violation of the commutative rule in CP [54], [42]. Violation of the sure thing principle (STP) and the total probability law (TPL) is observed repeatedly in practical tasks such as the "Two-stage gambling game" and "prisoner's dilemma" [53]. Because the TPL plays a key role in some widely used models, such as BNs, applying these models in human decision-making is a challenging issue. In the following, an example of the violation of TPL in the BN structure is presented.

### B. Violation of TPL in BN Structure

A BN is a directed acyclic graph containing some nodes, directional arcs, and a set of probability tables for each node. This structure is widely used in various fields, such as engineering, economics, social sciences, medicine, and many more [54][8]. The inferences in this method are based on CP and specifically Baye's rule for two events including $X$ and $Y$:

$$Pr(X|Y) = Pr(Y|X) Pr(X)/Pr(Y). \qquad (1)$$

Bayes Nets are a compact way to represent the joint distribution of a set of random variables, represented by nodes in BN. Each node $(X_i)$ has a conditional probability distribution $Pr(X_i|parents(X_i))$ that quantifies the effect of the parents on the node. So, the probability of a specific world state is obtained from a Bayes net as follows:

$$Pr(X_1, X_2, \dots, X_n) = \prod_{i=1}^{n} Pr(X_i|parents(X_i)). \qquad (2)$$

Other probabilities can be found from the CPT. Let us present a BN for modeling the PD task. In this task, there are two prisoners (A and B) in separated isolation cells. Each prisoner can decide to be silent (Cooperate) or betray (Defect). The payoff matrix for the PD is shown in Table I. When (B) wants to make a decision, he/she is informed that A has chosen to defect (D), cooperate (C), or has no information about A. Fig. 1 presents a classical BN for modeling the PD task based on experimental results presented in [53]. In the classical view, when prisoner B has no information about A (Unknown situation), the information for each option is interpreted separately by TPL, as shown in (3):

$$Pr(B = D) = Pr(D, D) + Pr(C, D)$$
$$= Pr(B = D|A = D) Pr(A = D)$$
$$\quad + Pr(B = D|A = C) Pr(A = C). \qquad (3)$$

However, these results do not match the experimental results, as shown in Table II.

### C. The Quantum Probability

The classical probability theory has been classified as classical measure theory [55]. Similarly, the quantum probability theory emerges from a non-commutative measure theory, generally studied in a suitable Hilbert space [56][57]. In the general form of a Hilbert space, we deal with the amplitude and the phase in the complex vector space. In QIT, the equivalent of a classic bit is a quantum bit or qubit. The spin, as a famous example of a qubit, can be represented in a two dimensional complex Hilbert space, where $|0\rangle = (1 \ 0)^T$ and $|1\rangle = (0 \ 1)^T$ are called eigenstates of the system (here the qubit). These eigenstates represent the vertical and horizontal polarization of a photon. In a classical system, a bit would have to be in one state or the other. The most fundamental property of quantum mechanics is that a qubit is in the state $|0\rangle$ and $|1\rangle$ simultaneously. However, it will collapse into one of the eigenstates when we measure the system. Without any measurement (under uncertainty), the state of a qubit is identified by a superposition of eigenstates as:

$$|\psi\rangle = c_1 e^{i\theta_1}|0\rangle + c_2 e^{i\theta_2}|1\rangle = (c_1 e^{i\theta_1} \ c_2 e^{i\theta_2})^T, \qquad (4)$$

where $c_n e^{i\theta_n} = \sqrt{Pr(|n-1\rangle)} \, e^{i\theta_n}$ coefficients are time-dependent, complex numbers represent by amplitude and phase. When a measurement occurs, the state function of the system collapses into one of the eigenstates $|0\rangle$ or $|1\rangle$. $|c_1|^2$, and $|c_2|^2$ are the probability of finding the system in $|0\rangle$ and $|1\rangle$, respectively, and $|c_1|^2+|c_2|^2 = 1$. For each qubit, an initial state vector $|\psi\rangle \in H$ is defined according to the initial conditions and our knowledge about the system. For each event $|i\rangle$, a projection operator $P_i = |i\rangle\langle i|$ is assigned to find $c_{i+1}$ and reduce the state vector $|\psi\rangle$ into eigenstate $|i\rangle$. The probability of finding the system in the state $|i\rangle$ is equal to the squared module of the corresponding projection value, as shown in (5):

$$Pr(|i\rangle) = |P_i|\psi\rangle|^2 = (P_i|\psi\rangle)^\dagger (P_i|\psi\rangle). \qquad (5)$$

To compute the length of a vector in this structure, generalization of the dot product in the complex domain is used [58]. Now we are ready to apply QP to investigate some decision-making problems.

## III. The proposed predictive entangled quantum-like Bayesian network

In this section, we try to present a realistic predictive model of human decision-making as well as estimate the interference term in the QBN structure. In the real world, we live as part of a complex multi-agent 'human' society, rather than in isolation.

TABLE I
THE PAYOFF MATRIX FOR THE PRISONER'S DILEMMA GAME

|  |  | Prisoner B | |
|---|---|---|---|
|  |  | C | D |
| Prisoner A | C | A: 1 year<br>B: 1 year | A: 10 year<br>B: 0 year |
|  | D | A: 0 year<br>B: 10 year | A: 5 year<br>B: 5 year |

|  | $Pr(\mathbf{B=C})$ | $Pr(\mathbf{B=D})$ |
|---|---|---|
| A=C | 0.16 | 0.84 |
| A=D | 0.03 | 0.97 |

$Pr(A = C) = 0.5$
$Pr(A = D) = 0.5$

Fig. 1. Classical BN for PD task. Each prisoner can be considered as one node. The first prisoner (A), can select cooperation (C) or defect (D) with equal probability (50%). The probability table of decisions made by the second prisoner (B) in the absence of uncertainty, based on empirical data presented in [53], is shown in this classical BN.

TABLE II
COMPARING THE PREDICTION OF CLASSICAL BN ACCORDING TO CP RULE (TPL), AND EXPERIMENTAL RESULTS IN [53]

|  | Predicted Probability by Classical BN | Experimental Results [53] |
|---|---|---|
| $Pr(B = D)$ | 0.5*0.97+0.5* 84=**0.905** | 0.63 |
| $Pr(B = C)$ | 0.5*0.03+0.5* 0.16=**0.095** | 0.37 |

We simulate a society as a composite quantum system in which DMs with different initial mental conditions are considered particles in the different superposition states. (Fig. 2) The novelty of our approach is considering the effect of other agents on the selection behavior of each DM by entanglement measures inspired by QIT. We introduce a predictive entangled quantum-like BN structure (PEQBN) for modeling the human decision-making process. In quantum physics, entanglement considers long-range effects due to unknown sources in the dynamical behavior of the system. We model binary decision-making experiments in a QBN structure in which a directed graph and conditional probability tables are defined, as shown in Fig. 3. The wave function in this figure is $\psi_x = \sqrt{Pr_x} e^{i\theta_x}$, where $Pr_x$ is the corresponding probability in classical BN. This structure, including graphs and tables, is used in all QBN models. However, the nodes in our model can be simulated as entangled q-bits (Dashed line in Fig. 3). In the proposed model, entangled nodes mean that each decision-maker/node is considered as a part of a whole system. So we consider indirect relations in the system that cannot be modeled as classical arcs. However, these relations affect environmental conditions and, therefore, human decisions. In the proposed model, similar to other QBN models, dynamical evolution about selecting choices under uncertainty can be assumed as overlapped wave functions. The interference of these waves is the function of environmental/initial conditions. The distinguishing feature of the presented model is relating interference terms to the entanglement property of a system. Let us explain this approach in detail by considering a QBN with two nodes.

As mentioned in Section II, each quantum event in QP is characterized by a state vector in Hilbert space. In a binary problem, two available choices for each node, including $|a_1\rangle = (1\ \ 0)^T$ and $|a_2\rangle = (0\ \ 1)^T$ in Hilbert space $H_1$ are defined. Similarly $|b_1\rangle = (1\ \ 0)^T$ and $|b_2\rangle = (0\ \ 1)^T$ in Hilbert space $H_2$ are defined too. Node $A$ (parent node) can exist in the following superposition state:

$$|\psi_A\rangle = \kappa_1 e^{i\gamma_1}|a_1\rangle + \sqrt{1-\kappa_1^2}e^{i\gamma_2}|a_2\rangle = \left(\kappa_1\ e^{i\gamma_1}\quad \sqrt{1-\kappa_1^2}\ e^{i\gamma_2}\right)^T. \qquad (6)$$

If the uncertainty about node A is eliminated by doing a measurement on A, the wave function $|\psi_A\rangle$ will be collapsed into one of the eigenstates $|a_1\rangle$ or $|a_2\rangle$. For extending the decision problem to a 4-dimensions, the composite Hilbert space $H = H_1 \otimes H_2$ and four basis vectors are obtained by the tensor product of $H_1$ and $H_2$:

$|a_1 b_1\rangle = |a_1\rangle \otimes |b_1\rangle = (1\ \ 0\ \ 0\ \ 0)^T,$
$|a_1 b_2\rangle = |a_1\rangle \otimes |b_2\rangle = (0\ \ 1\ \ 0\ \ 0)^T,$
$|a_2 b_1\rangle = |a_2\rangle \otimes |b_1\rangle = (0\ \ 0\ \ 1\ \ 0)^T,$
$|a_2 b_2\rangle = |a_2\rangle \otimes |b_2\rangle = (0\ \ 0\ \ 0\ \ 1)^T. \qquad (7)$

The Hilbert space H represents all possible results after a measurement. For example, $|a_1 b_1\rangle$ is related to the state that the decision in node $A$ results $a_1$ and the decision in node $B$ is $b_1$. So until we are not aware of making a decision (doing a measurement) in nodes $A$ and $B$, the following superposition mode is formed:

$$|\psi_{AB}\rangle = c_1 e^{i\theta_1}|a_1 b_1\rangle + c_2 e^{i\theta_2}|a_1 b_2\rangle + c_3 e^{i\theta_3}|a_2 b_1\rangle + c_4 e^{i\theta_4}|a_2 b_2\rangle = (c_1 e^{i\theta_1}\quad c_2 e^{i\theta_2}\quad c_3 e^{i\theta_3}\quad c_4 e^{i\theta_4})^T, \qquad (8)$$

where $c_i$ and $\theta_i$ are defined in Table III and $\sum_{i=1}^{4}|c_i|^2 = 1$. For example, for estimating $\Pr(B = b_1)$ without measuring A, the first step is applying the projector $P_{B=b}$, equal to outer product $|b_1\rangle\langle b_1|$, on the wave function $\psi_{AB}$. So, according to (5), and using the inner product in the complex domain [58]:

$$\Pr(B = b_1) = |P_{B=b_1}|\psi_{AB}\rangle|^2 = |\langle b_1|\psi_{AB}\rangle|^2 \qquad (9)$$
$$= |\langle b_1|a_1\rangle\langle a_1|\psi_{AB}\rangle + \langle b_1|a_2\rangle\langle a_2|\psi_{AB}\rangle|^2$$
$$= c_1^2 + c_3^2 + (|\langle b_1|a_1\rangle||\langle a_1|\psi_{AB}\rangle||\langle \psi_{AB}|a_2\rangle||\langle a_2|b_1\rangle|cos\theta).$$

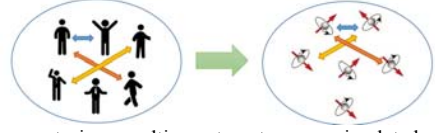

Fig. 2. The agents in a multi-agent system are simulated as particles in a composite quantum system. Each DM/particle has a particular initial mental/superposition condition before reaching the final decision/ measuring. So inspired by quantum computing each DM consider as a part of the whole system, and the unknown relations in the society are modeled as long-range stochastic effects by the entanglement.

The last term in this equation is the interference term, which adds a long-range stochastic effect on the problem. If we have eliminated uncertainty about the first node by observing it, the third term will equal zero. In this case, $\Pr(B = b_1)$ becomes equal to $c_1^2$ or $c_3^2$ if observing A results $|a_1\rangle$ or $|a_2\rangle$ respectively. While, if we do not have any information about node $A$, we need to estimate the interference term in the above equation. The interference term is related to the initial condition, including amplitudes and phases in the superposition state. In this method, we consider that the initial phase is related to the effect of society and other agents on DM. So we estimate the key part of the interference term $(cos(\theta))$ by measuring the entanglement of the system practically. We assume that the interference term of such systems is related to the entanglement of formation $E(\rho)$ [58]:

$$cos(\theta) = -E(\rho), \qquad (10)$$

where $\rho$ is a density matrix defined in Appendix 1. In QIT, there are many criteria called measures and witnesses to determine a system is entangled or not and estimate the value of entanglement. Two famous entanglement measures are concurrence $C(\rho)$ and Shanon entropy $E_{Sh}(\rho)$. As seen in [58], we can use Shanon entropy for estimating $E(\rho)$. The definition of $E_{Sh}(\rho)$ and the relation of this entropy to $C(\rho)$ in the binary problem is presented in (11) [58]:

$$E(\rho) = E_{Sh} = -m\log_2 m - (1-m)\log_2(1-m),$$
$$\text{where } m = \left(\frac{1+\sqrt{1-C(\rho)^2}}{2}\right). \qquad (11)$$

Unfortunately, there is no procedure to find initial phases in the density matrix of the social system and estimate $E(\rho)$ directly, as presented in QIT (Appendix 1). So we introduce a quantum-like witness $(qW)$ in the observed situation/classical probability. Then we apply $(qW)$ to estimate the value of $C(\rho)$ in an uncertain case.

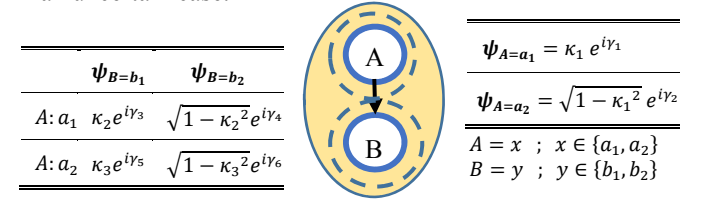

| | $\psi_{B=b_1}$ | $\psi_{B=b_2}$ |
|---|---|---|
| $A: a_1$ | $\kappa_2 e^{i\gamma_3}$ | $\sqrt{1-\kappa_2^2}e^{i\gamma_4}$ |
| $A: a_2$ | $\kappa_3 e^{i\gamma_5}$ | $\sqrt{1-\kappa_3^2}e^{i\gamma_6}$ |

| |
|---|
| $\psi_{A=a_1} = \kappa_1\ e^{i\gamma_1}$ |
| $\psi_{A=a_2} = \sqrt{1-\kappa_1^2}\ e^{i\gamma_2}$ |
| $A = x\ ;\ x \in \{a_1, a_2\}$ |
| $B = y\ ;\ y \in \{b_1, b_2\}$ |

Fig 3. The general structure of entangled BNs for two nodes. Each node/agent is considered as a part of the whole multi-agent system (yellow oval). We consider entanglement due to unknown sources such as cooperation between agents. Also, similar to other models of quantum-like BNs, probabilities in BNs are replaced by suitable complex wave functions in these tables.

TABLE III
WAVE FUNCTIONS IN PEQBN MODEL PRESENTED IN FIG 3

| A | B | $\psi(A,B)$ |
|---|---|---|
| $a_1$ | $b_1$ | $\psi(a_1,b_1) = \kappa_1 e^{i\gamma_1} \kappa_2 e^{i\gamma_3} = c_1 e^{i\theta_1}$ |
| $a_1$ | $b_2$ | $\psi(a_1,b_2) = \kappa_1 e^{i\gamma_1}\sqrt{1-\kappa_2^2}\ e^{i\gamma_4} = c_2 e^{i\theta_2}$ |
| $a_2$ | $b_1$ | $\psi(a_2,b_1) = \sqrt{1-\kappa_1^2}\ e^{i\gamma_2}\kappa_3\ e^{i\gamma_5} = c_3 e^{i\theta_3}$ |
| $a_2$ | $b_2$ | $\psi(a_2,b_2) = \sqrt{1-\kappa_1^2}\ e^{i\gamma_2}\sqrt{1-\kappa_3^2}\ e^{i\gamma_6} = c_4 e^{i\theta_4}$ |

### A. Introducing a Quantum-Like witness

Let us consider the simplest structure of the proposed quantum-like BN in a binary decision-making problem in Fig. 3. The probability of finding the system in $A = a_i$ and $B = b_j$ by observing their nodes is $Pr(a_i,b_j) = |\psi(a_i,b_j)|^2$. Let us assume the state of the first node is $a_1$. In similar terms of profit and loss, the difference between the probability of choosing $b_1$ and $b_2$ is not necessarily the same in all communities. So we define the following variable:

$$\Delta b_{a_i} = \Pr(a_i,b_1) - \Pr(a_i,b_2). \quad (12)$$

This variable depends on the environmental conditions in each society (the degree of entanglement in the system). The probability of finding a system in the $B = b_1$ in the classical view is obtained by $Pr(a_1,b_1) + Pr(a_2,b_1)$. But we use a weighted summation of these probabilities and introduce a quantum-like witness ($qW$) as follows:

$$qW = \cos^{-1}\{\delta(\Pr(a_1,b_1)*\Delta b_{a_1} + \Pr(a_2,b_1)*\Delta b_{a_2})\}, \quad (13)$$

where

$$\delta = 1/\sqrt{((\Pr(a_1,b_1))^2 + (\Pr(a_2,b_1))^2)(\Delta b_{a_1}^2 + \Delta b_{a_2}^2)}.$$

The important feature of the proposed quantum-like witness is its independence from phase parameters. We propose a relationship between the introduced witness and $C(\rho)$ by using an optimization algorithm. This procedure is meaningful because optimization is used in the definition of most quantum measures and witnesses. We find the following relationship between the proposed $qW$ and concurrence measure $C(\rho)$:

$$C(\rho) = \quad (14)$$

$$\sqrt{z_0 + \sum_{i=1}^{2}(-1)^i(z_{2i-1}\cos(i*12\ qW) + z_{2i}\sin(i*12\ qW))}.$$

where z-coefficients are constant numbers obtained using an optimization algorithm as: $z_0 = 0.216$; $z_1 = 0.073$ ; $z_2 = 0.126$; $z_3 = -0.140$; $z_4 = 0.069$.

This relation guarantees that $0 < C(\rho) < 1$, which is compatible with the acceptable values for this measure. The constant coefficients in this relation are obtained by applying the optimization algorithm on the experimental data of 9 PD and two-stage gambling game tasks presented in Section IV.A-B. Through this procedure, concurrence, Shannon entropy, and the interference term are estimated as presented in Table IV.

### B. Numerical Example of Applying PEQBN

Here, we apply our approach to the PD task. As mentioned in Section II.B, in this game there are two prisoners in the separated cells and unaware of another decision. They are considered entangled nodes/agents in our method due to unknown sources such as previous cooperation or cognition of each other's spirits. So, each prisoner is considered part of the whole system (yellow oval in Fig. 3) instead of an independent agent. We define two vectors for states of A and two vectors for states of B in the Hilbert space $H_1$ and $H_2$, respectively:

TABLE IV
PSEUDO-CODE OF PEQBN MODEL FOR BINARY DECISION-MAKING SCENARIO WITH TWO NODES INCLUDING A AND B

**Step 1:** Two vectors for states of each binary node are presented as: $|a_1\rangle = (1\ \ 0)^T$ and $|a_2\rangle = (0\ \ 1)^T$ in Hilbert space $H_1$ and $|b_1\rangle = (1\ \ 0)^T$ and $|b_2\rangle = (0\ \ 1)^T$ in $H_2$.

**Step 2:** The composite Hilbert space $H = H_1 \otimes H_2$ and four basis vectors are obtained by the tensor product.

**Step 3.** Superposition mode $|\psi_{AB}\rangle$ is formed as presented in (8).

**Step 4:** Quantum-like witness ($qW$) is calculated based on (13).

**Step 5:** Concurrence measure $C(\rho)$ is calculated based on (14).

**Step 6:** Shannon entropy $E_{Sh}(\rho)$ is calculated based on (11).

**Step 7.** For estimating $\Pr(B = b_1)$, the projector $P_{B=b_1}$ is applied on $|\psi_{AB}\rangle$ as shown in (9) and $\cos(\theta)$ in (9) is considered equal to $-E_{Sh}(\rho)$.

$|A = D\rangle = (1\ \ 0)^T$ ; $|A = C\rangle = (0\ \ 1)^T$,
$|B = D\rangle = (1\ \ 0)^T$ ; $|B = C\rangle = (0\ \ 1)^T$.

The superposition state is obtained as follows based on (8) and the experimental results in [53]:

$$|\psi_{AB}\rangle = 0.696 e^{i\theta_1}|DD\rangle + 0.123 e^{i\theta_2}|DC\rangle$$
$$+0.648 e^{i\theta_3}|CD\rangle + 0.282 e^{i\theta_4}|CC\rangle$$
$$= (0.696 e^{i\theta_1}\ \ 0.123 e^{i\theta_2}\ \ 0.648 e^{i\theta_3}\ \ 0.282 e^{i\theta_4})^T.$$

$|\psi_{AB}\rangle$ is considered as the initial state of the second prisoner's mind in the dynamic evolution of the decision-making process. The phase parameters in $|\psi_{AB}\rangle$ are related to the environmental condition of the whole system, which is assumed related to unknown relation between two prisoners in our approach. Then, Δ variables are formed according to (12):

$\Delta b_{A=D} = \Pr(D,D) - \Pr(D,C) = 0.47$,
$\Delta b_{A=C} = \Pr(C,D) - \Pr(C,C) = 0.34$.

Moreover, $qW$, $C(\rho)$, and $E_S(\rho)$ are calculated for estimating $\Pr(B = D)$ by applying (13), (14), (11), and (10), respectively:

$$qW = \cos^{-1}\left\{\frac{0.485*0.47 + 0.42*0.34}{\sqrt{(0.485^2 + 0.42^2)(0.47^2 + 0.34^2)}}\right\} = 0.087,$$

$C(\rho) = 0.447, E_{Sh}(\rho) = 0.298, \cos(\theta) = -E(\rho) = -0.298.$

So the probability of the second prisoner decides to defect is calculates by (9):

$Pr\ (B = D\ |\ A\ \text{is}\ Unknown) = 63.5\%$,
$Pr\ (B = C\ |\ A\ \text{is}\ Unknown) = 100 - 63.5 = 36.5\%$.

As can be seen, the obtained $\Pr(B = D) = 63.5\%$ by PEQBN is in good agreement with the empirical result (63%) [53] (Table V) compared to the predicted $\Pr(B = D) = 90.5\%$ by CBN model (Table II). The comprehensive evaluation of this model on 22 experiments is reported in detail in Section IV. These evaluations show a better performance of PEQBN compare to CBN and three quantum-like decision models.

## IV. Evaluation of PEQBN on modeling human decision

Now we are ready to compare our proposed PEQBN to CBN as well as three recent quantum-like decision-making models [28][29][42] to provide a fair comparison. We have compared our results with the outcomes of other models for three tasks. These tasks are the prisoner's dilemma, two stages gambling game, and document relevance judgment experiment.

## A. Task 1. Evaluation of PEQBN on Modeling Prisoners Dilemma

In the first evaluation task, the PD game, which is introduced in Section II.B, is modeled by the proposed PEQBN, CBN, QDT [29], and two different QBN methods [28][42]. Reported results by four studies in the literature on this task are shown in Table V. The absolute errors between empirical results and predicted probabilities are shown in Table VI and Fig. 4. In Table VI, also the root-mean-square error (RMSE) for each model is presented. According to these comparisons, PEQBN wins first place among all the compared algorithms by achieving RMSE = 4.51. Two other QBN models [28][42] by applying the heuristic function and Deng entropy, QDT[29], which is considered static interference term, and CBN without considering interference term are ranked second to fifth, respectively. So as shown in Fig 4, the proposed model results are in good agreement with the experimental data. The reduction of error between the predicted and empirical results in the proposed method can be related to the existence of cooperative relationships between agents/nodes in this social system, which is modeled with the idea of entanglement in the PEQBN model.

## B. Task 2. Evaluation of PEQBN on Modeling Two-Stage Gambling Game

The second task is about the modeling of a two-stage gambling game. In this game, participants are asked to "play" or "not play" in a two-stage gamble. In each stage, participants have a 50% chance of winning 200$ or losing 100$. For the second stage of this task, the player is informed that she/he has won or lost in the first gamble, or has no information about the first round (Unknown). The experimental results used in this task are presented in Table VII. The deviation between the empirical and predicted probability of choosing "play" in the second round, while the outcome of the first round is not informed, is presented in Table VIII and Fig 5. Similar to the previous experiment, the predicted probabilities are obtained by the PEQBN model as well as classical BN and three recent quantum-like decision models, including QDT [29] and two different QBN models presented in [28], and [42]. Also, RMSE values for each model have been presented in Table VIII for better comparison.

Fig. 5 and obtained RMSE value equal to 0.06 for PEQBN show a significant improvement in modeling human selection behavior compare to CBN, QDT [29], and two different QBN methods [28][42]. According to these comparisons, PEQBN wins first place among all the compared algorithms again. QBN model [28], which is considered heuristic function, and QDT [29] by considering static interference term are ranked second to third respectively. QBN based on Deng entropy [42], and CBN without considering interference term achieve jointly the fourth rank and presented similar performance. These results confirm that two stages in a gambling game find the best description when we look at them as entangled entities. Winning or losing in the first stage changes the initial conditions (decision maker's mental state) and will affect choosing "play" for the second stage.

## C. Task 3. Evaluation of PEQBN on Document Relevance Judgment

In previous tasks, we evaluate PEQBN on the empirical data applied in the optimization process of finding (14). If this relation is valid, it should provide acceptable results for other databases. Here, PEQBN is applied on a fresh data set, about document relevance judgment, for implementing a comprehensive evaluation. In this experiment, presented by Wang et al. [52], 15 separate sessions are defined. In each session, participants must judge two documents titled A and B concerning a specific query. For example, in the first session, the title of A and B are *statics* and *distribution*, and the title of the query is *probability*.

TABLE V
THE PROBABILITY OF A SECOND PLAYER CHOOSING TO DEFECT UNDER SEVERAL CONDITIONS ABOUT THE DECISION OF THE FIRST PLAYER

|  | Known to | | Unknown | |
|---|---|---|---|---|
|  | Defect | Cooperate | Experimental | CP (TPL) |
| Shafir & Tversky [53] | 0.97 | 0.84 | 0.63 | 0.905 |
| Li & Taplini [59] | 0.82 | 0.77 | 0.72 | 0.795 |
| Busemeyer et al. [60] | 0.91 | 0.84 | 0.66 | 0.875 |
| Hristova & Grinberg [61] | 0.97 | 0.93 | 0.88 | 0.950 |

TABLE VI
COMPARING ABSOLUTE ERRORS BETWEEN EXPERIMENTAL AND PREDICTED PROBABILITY OF PR(B = DEFECT) UNDER UNCERTAINTY IN THE PD GAME FOR FIVE MODELS INCLUDING CBN, QDT [29], QBN [28], QBN[42] AND THIS STUDY

| PD Game | CBN | QDT[29] | QBN[28] | QBN[42] | PEQBN |
|---|---|---|---|---|---|
| **Shafir & Tversky [53]** | 27.50 | 2.50 | 1.08 | -11.16 | **0.54** |
| **Li & Taplini [59]** | 7.50 | -17.50 | **-0.78** | -14.07 | -6.98 |
| **Busemeyer et al. [60]** | 21.50 | **-3.50** | 13.95 | -5.31 | 5.58 |
| **Hristova & Grinberg[61]** | 7.00 | -18.00 | 2.64 | 2.45 | **1.14** |
| **RMSE** | 18.19 | 12.74 | 7.13 | 9.44 | **4.51** |

TABLE VII
THE PROBABILITY OF THE PLAYER CHOOSING TO PLAY AGAIN, BASED ON THE RESULTS OF THE FIRST STEP

|  | Known to | | Unknown | |
|---|---|---|---|---|
|  | Win | Lose | Experimental | CP(TPL) |
| Tversky and Shafir [21] | 0.69 | 0.58 | 0.37 | 0.635 |
| Kuhberger et al.[62] | 0.72 | 0.47 | 0.48 | 0.590 |
| Lambdin and Burdsal [63] | 0.63 | 0.45 | 0.41 | 0.540 |

TABLE VIII
COMPARING ABSOLUTE ERRORS BETWEEN THE EMPIRICAL AND PREDICTED PROBABILITY OF THE PLAYER CHOOSING PLAY IN THE SECOND ROUND OF GAMBLING GAME UNDER UNCERTAINTY FOR FIVE MODELS INCLUDING CBN, QDT [29], QBN[28], QBN[42], AND THIS STUDY.

| Gambling Game | CBN | QDT[29] | QBN[28] | QBN[42] | PEQBN |
|---|---|---|---|---|---|
| **Tversky and Shafir [21]** | 26.50 | 1.50 | -0.59 | 26.28 | **0.04** |
| **Kuhberger et al. [62]** | 11.50 | -13.50 | -7.82 | 11.87 | **-0.07** |
| **Lambdin and Burdsal [63]** | 13.00 | -12.00 | 4.93 | 13.11 | **-0.05** |
| **RMSE** | 18.29 | 10.46 | 5.35 | 18.29 | **0.06** |

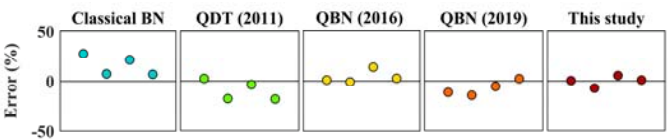

Fig 4. Error values between experimental results, presented in Table V, and predicted values of $\Pr(B = D)$ in the PD game. The predicted results are obtained by five different models including classical BN, three recently published quantum-like decision models [28][29][42], and this study.

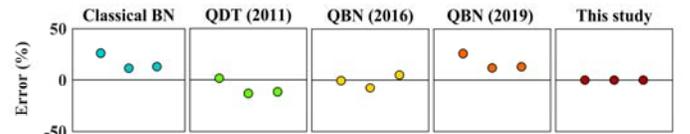

Fig 5. Error values between experimental results, presented in Table VII, and the predicted probability of the player choosing to play in the second round of a two-stage gambling game under uncertainty. The predicted results are obtained by five different models including classical BN, three recently published quantum-like decision models [28][29][42], and this study.

The participants are randomly divided into two groups. The first group should judge the B and A in the BA order on the relevance to the *probability* query. But the second group must judge about A without decision-making on B. The title of A and B documents and the subject of queries are shown in Table IX. The empirical results about judging on A by two different groups are demonstrated in Table X. We have applied PEQBN, CBN, QDT [29], and two different QBN [28][42] on empirical results in this table. The deviations between empirical and predicted probabilities of relating A to the specified query, without judging on B, are presented in Table XI and demonstrated in Fig. 6. Similar to other evaluation tasks, the RMSE value for each model is also presented in Table XI. As shown in Fig. 6 and Table XI, the PEQBN model shows the best agreement to empirical results as compared to CBN based on TPL and three recent quantum-like predictive decision models [28] [29] [42]. According to a TPL in the second group:

$$Pr(A \text{ is relevant}) = \qquad (15)$$
$$Pr(A \text{ is relevant} | B \text{ is relevant}) * Pr(B \text{ is relevant}) +$$
$$Pr(A \text{ is relevant} | B \text{ is} \sim relevant) * Pr(B \text{ is} \sim relevant).$$

But based on the PEQBN model, judging about A and B are two entangled processes. Measuring B (Judging about B) changes the initial conditions (decision maker's mental state) and will affect judging about A. This effect is not modeled, by classical cause and effect arcs, in the BN structure. PEQBN models this effect by entanglement concept through the interference term. Therefore, the obtained results from the second group, which judges only about A, differ from the calculated value predicted by TPL. Similar to previous tasks, PEQBN wins first place among all the compared algorithms. But, in this task, which participants do not receive any benefit or loss according to their choices, other quantum-like methods are no more successful than the classical method. Thus, CBN and QDT [29], by considering values equal to 0 and 0.25 for interference terms, are ranked second to third. And two previous quantum-like QBN [28] and [42] are ranked fourth to fifth by estimating interference terms based on heuristics value and Deng entropy. Since data in this task is not used in proposing our approach, achieving the first rank by PEQBN in this experiment is more valuable compared to previous tasks.

*D. Conceptual and numerical comparison with classical BN and three predictive quantum-like decision-making models*

This study proposed the predictive quantum-like BN for modeling human selection behavior, while most of the quantum-like decision-making methods are purely descriptive. Similar to other quantum-like decision theories, the evolution of different decisions are simulated as overlapping wave functions under uncertainty. Also, CP is replaced by QP in computational steps. Like previous quantum-like BNs, we consider interconnected nodes and model cause-and-effect relationships by arcs. But, we improve the previous methods by presenting innovations as follows:

(1) The main novelty in our approach is considering the effect of society on each agent by using two well-known entanglement measures and the proposed quantum-like witness. Entanglement models unknown long-range relationships between particles in a composite quantum system. In such systems, each particle is modeled as part of the whole system instead of an independent particle. So, it can be very inspiring for modeling unknown complex relationships between human agents in a multi-agent human system. Hence in our model, each human is considered as part of the whole society instead of an independent agent.

(2) We estimated the interference term by the social and physical concept confirmed by some neurophysiological evidence [51] instead of considering constant values equal to 0 and 0.25 in CBN and QDT [29] or heuristic function in [28].

TABLE IX
THE TITLE OF DOCUMENTS A AND B AND THE QUERY TERM ON WHICH DOCUMENTS A AND B ARE JUDGED [52]

| Query | Title of A | Title of B |
|---|---|---|
| Probability | Statistics | Distribution |
| Parkinsonism | Parkinson's and Alzheimer's diseases | Alzheimer's disease |
| Albert Einstein | Isaac Newton | Theory of relativity |
| The Spring Festival | Jiaozi | Grabbing tickets for the Spring Festival |
| Innovation-driven | Sharing economy | New open economic system |
| Machine learning | Artificial intelligence | Pattern recognition |
| American president | Barack Obama | Obama's wife |
| Semantic Web | Theology | Ontology |
| Computer | Boolean algebra | Turing machine |
| Religion | Politics | Culture |
| Kung Fu Panda | The Giant panda | Chinese martial arts |
| Air pollution | Haze | Spiritual therapy for smog |
| Transgene | Hybrid | Garden roses |
| Chinese literature | Classical prose | Chinese cuisine |
| Mo Yan | Nobel Prize | Red Sorghum (novel) |

TABLE X
THE EMPIRICAL RESULTS PRESENTED IN [52] OF PARTICIPANT'S JUDGMENTS ON THE SUBJECT OF DOCUMENTS A AND B.

| | First Group | | | Second Group |
|---|---|---|---|---|
| | Pr (B is relevant) | Pr (A is relevant \| B is relevant) | Pr (A is relevant \| B is not relevant) | Pr (A is relevant) |
| Probability | 0.58 | 0.89 | 0.62 | 0.83 |
| Parkinsonism | 0.84 | 0.46 | 0.6 | 0.3 |
| Albert Einstein | 0.19 | 1 | 0.92 | 0.9 |
| The Spring Festival | 0.52 | 0.81 | 0.8 | 0.6 |
| Innovation driven | 0.06 | 0.5 | 0.45 | 0.33 |
| Machine learning | 0.68 | 0.95 | 1 | 0.9 |
| American president | 1 | 0.52 | 0 | 0.37 |
| Semantic Web | 0.03 | 1 | 0.4 | 0.2 |
| Computer | 0.42 | 1 | 0.72 | 0.7 |
| Religion | 0.26 | 0.88 | 0.78 | 0.53 |
| Kung Fu Panda | 0.42 | 0.76 | 0.29 | 0.5 |
| Air pollution | 0.81 | 0.92 | 0.83 | 0.83 |
| Transgene | 0.39 | 0.5 | 0.11 | 0.07 |
| Chinese literature | 0.87 | 0.26 | 0.5 | 0.4 |
| Mo Yan | 0.42 | 0.94 | 0.93 | 0.97 |

TABLE XI
COMPARING ABSOLUTE ERRORS BETWEEN EXPERIMENTAL AND PREDICTED PROBABILITY OF DOCUMENT A IS RELEVANT TO ITS QUERY TERM IF DOCUMENT B HAS NOT BEEN JUDGED FOR FIVE MODELS INCLUDING CBN, QDT [29], QBN[28], QBN[42], AND THIS STUDY.

| Query | CBN | QDT [29] | QBN[28] | QBN[42] | PEQBN |
|---|---|---|---|---|---|
| 1 | **-5.340** | -30.340 | -10.621 | -7.213 | -19.179 |
| 2 | 18.240 | -6.760 | 5.471 | 22.700 | **-5.205** |
| 3 | 3.520 | -21.480 | -13.932 | -5.976 | **-0.938** |
| 4 | 20.520 | **-4.480** | 31.438 | 19.800 | 18.773 |
| 5 | 12.300 | -12.700 | 10.320 | 14.289 | **5.284** |
| 6 | 6.600 | -18.400 | -8.362 | -3.177 | **1.472** |
| 7 | 15.000 | **-10.000** | 15.000 | 15.000 | 15.000 |
| 8 | 21.800 | **-3.200** | 5.786 | 63.954 | 13.560 |
| 9 | 13.760 | -11.240 | -41.661 | 8.423 | **0.122** |
| 10 | 27.600 | **2.600** | 10.026 | 24.421 | 10.288 |
| 11 | 4.850 | -20.150 | 16.622 | **3.221** | -23.711 |
| 12 | 7.290 | -17.710 | 13.426 | 7.092 | **2.322** |
| 13 | 19.210 | **-5.790** | 20.659 | 19.214 | 6.673 |
| 14 | -10.880 | -35.880 | -26.024 | **-3.183** | -15.188 |
| 15 | -3.450 | -28.450 | **0.277** | -2.515 | -4.705 |
| RMSE | 14.640 | 18.330 | 18.420 | 21.050 | **11.990** |

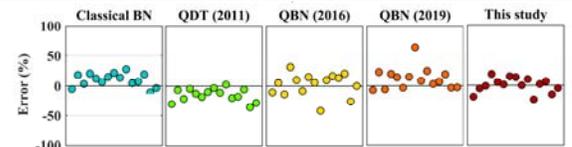

Fig. 6. Errors between observed experimental results and predicted probability of "document A is relevant to its query term" if document B has not been judged. Predicted results are obtained by five models including CBN, three recently quantum-like decision models [28][29][42], and this study.

Although Huang et al. [42] tried to present a logical justification for estimating interference effects, their method has some drawbacks because of the improper use of physical and mathematical functions. Let us apply their method, for example, on the PD task. They used Deng entropy for estimating the interference term to measure the uncertainty degree of the event $y \in \{C, D\}$ [42] as follows :

$$\cos(\theta) = -\text{Deng entropy} = \sum_y B_d(y) \log(B_d(y)/(2^{|n|} - 1)), \quad (16)$$

where $|n|$ is equal to the number of answers to a query [42]. $B_d(y)$, for the event $y$, is estimated by:

$$B_d(y) = \begin{cases} \left|\alpha_y + \frac{\alpha_y - \beta_y}{|\alpha_y + \beta_y - 1|}\right| & if \ |\alpha_y - 0.5| < |\beta_y - 0.5| \\ \left|\beta_y + \frac{\beta_y - \alpha_y}{|\beta_y + \alpha_y - 1|}\right| & if \ |\alpha_y - 0.5| \geq |\beta_y - 0.5| \end{cases},$$

$$\alpha_y = |\psi(D, y)|; \ \beta_y = |\psi(C, y)|. \quad (17)$$

It is noted that according to the definition of Deng entropy [43], $B_d(y)$ must be a probability-like function that varies between 0 and 1 and satisfy the following conditions:

$$B_d(\emptyset) = 0; \quad \sum_y B_d(y) = 1; \quad (18)$$

This limitation guarantee that the absolute value of Deng entropy, which is considered equal to $\cos(\theta)$ in [42], does not exceed one. But according to the definition of $B_d(y)$ in (17), this function can take a value greater than one which leads to assign values greater than one to the cosine function too. We plot $B_d(\text{Cooperate})$, which are presented by Huang et al. [42], for possible values between 0 to 1 of $Pr(B = C|A = D)$ and $Pr(B = C|A = C)$ in Fig. 7 (a). As shown in this figure, $B_d(\text{Cooperate})$ does not satisfy the mathematical constraint of introduced belief distance ($B_d$) in [43]. Therefore, it seems that the application of $B_d(y)$ in estimating the interference term is not meaningful in all cases. In Fig 7(b), we plot $\cos(\theta)$ for possible values between 0 to 1 of $Pr(B = C|A = D)$, and $Pr(B = C|A = C)$. As this figure demonstrates, the absolute values for $\cos(\theta)$ are not bounded to 1. The white area in Fig. 7(a) is related to situations in which $B_d(\text{Cooperate}) > 1$, and the application of Deng entropy is not meaningful. So one advantage of our approach compared to their method is that all functions in our method (like $\cos(\theta)$, and $E_{Sh}(\rho)$) are bounded to their physical and mathematical constraints. So, PEQBN can be used in all situations.

(3) Next advantage of our method compared to the method presented in [28], [29], and [42] is that we consider a continuous range between 0 and 1 for the interference term. CBN is not considered the interference term. QDT [29] is considered only one static value equal to 0.25, and QBN [28] considered piecewise function. Although QBN presented by Huang et al. [42] estimates the interference term continuously, this model does not generate meaningful value for all inputs. Besides, we have a quantum view in all parts of the quantum-like procedure.

(4) Finally, the absolute advantage of our approach is achieving the first rank in three evaluation tasks and presenting a better match between predicted probabilities and experimental data used in Section IV.A-C. Table XII shows the ranks of all compared models for the overall 22 experiments presented in this study. The average ranks for all compared models, presented in Table XII, confirm that PEQBN achieves first place by obtaining a minimum value equal to 2.09.

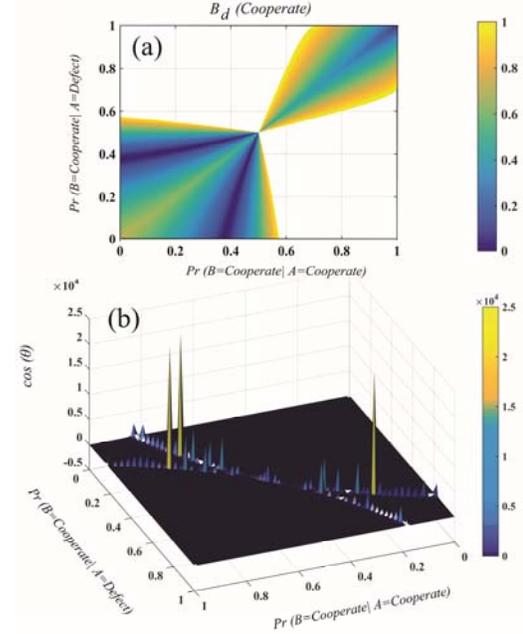

Fig.7. (a). $B_d(Cooperate)$ versus $Pr(B = C|A = D)$ and $Pr(B = C|A = C)$. The white area in this plot shows the range of $Pr(B = C|A = D)$ and $Pr(B = C|A = C)$ that lead to obtain $B_d(Cooperate) > 1$. (b). $\cos(\theta)$ in the interference term versus $Pr(B = C|A = D)$, and $Pr(B = C|A = C)$. The obtained values for cosine function in [42] are not bounded to |1|.

Other predictive quantum-like methods including two different QBN[28][42], and QDT[29] are ranked second to fourth by considering three distinct heuristic values, dynamic value based on Deng entropy, and static interference term, respectively. Based on the overall evaluations, the classical BN, without considering any interference term, performed worse than all the compared methods. Finally, RMSE values of the overall 22 experiments in this study have been presented in Fig. 8 for better comparison. PEQBN wins first place again by achieving an RMSE value equal to 10.09 among all the compared models.

It is noted that human selection behavior in realistic challenges of modern life, such as route selection, stock market investment, or sharing information in social networks, is highly dependent on the environmental situation and behavior of other agents. So PEQBN can be successfully applied in social systems by considering the entanglement concept and modeling the relations between agents similar to the stochastic long-range effect between particles.

V. CONCLUSION

Human selection behavior under uncertainty often violates the predictions of most human decision-making models that follow the principles of classical probability (CP) theory. The similarity between the violations of CP in decision-making and physical domains, which is addressed by using QP, has recently led to the idea of applying QP in decision-making tasks. QP has shown considerable ability to model human behavior by extending CP from real scalar to Hilbert vector space. Although quantum-like decision theories can explain most decision-making paradoxes, presenting a predictive quantum-like decision-making model is yet an unsolved challenge. Such a predictive model could be used in devising appropriate responses toward human-inclusive systems under crisis or for their control.

TABLE XII
THE RANK OF THE PROPOSED PEQBN MODEL IN THIS STUDY COMPARED TO FOUR MODELS INCLUDING CBN, QDT [29], QBN[28], AND QBN [42] FOR ALL 22 EXPERIMENTS IN SECTION IV

| Task | | CBN | QDT [29] | QBN[28] | QBN[42] | PEQBN |
|---|---|---|---|---|---|---|
| 1 | 1 (IV.A) | 5 | 3 | 2 | 4 | **1** |
| 2 | | 3 | 5 | **1** | 4 | 2 |
| 3 | | 5 | **1** | 4 | 2 | 3 |
| 4 | | 5 | 4 | 2 | 3 | **1** |
| 5 | 2 (IV.B) | 5 | 3 | 2 | 4 | **1** |
| 6 | | 3 | 5 | 2 | 4 | **1** |
| 7 | | 4 | 3 | 2 | 5 | **1** |
| 8 | 3 (IV.C) | **1** | 5 | 3 | 2 | 4 |
| 9 | | 4 | 3 | 2 | 5 | **1** |
| 10 | | 2 | 5 | 4 | 3 | **1** |
| 11 | | 4 | **1** | 5 | 3 | 2 |
| 12 | | 3 | 4 | 2 | 5 | **1** |
| 13 | | 3 | 5 | 4 | 2 | **1** |
| 14 | | 2 | **1** | 2 | 2 | 2 |
| 15 | | 4 | **1** | 2 | 5 | 3 |
| 16 | | 4 | 3 | 5 | 2 | **1** |
| 17 | | 5 | **1** | 2 | 4 | 3 |
| 18 | | 2 | 4 | 3 | **1** | 5 |
| 19 | | 3 | 5 | 4 | 2 | **1** |
| 20 | | 2 | 5 | 5 | **1** | 3 |
| 21 | | 3 | 5 | **1** | 2 | 4 |
| 22 | | 3 | 5 | **1** | 2 | 4 |
| Average Rank | | 3.41 | 3.50 | 2.68 | 3.05 | **2.09** |

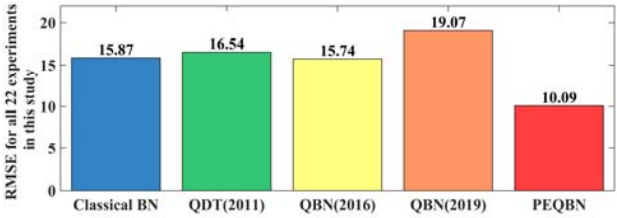

Fig. 8. The root-mean-square errors of all experiments in this study for five different models, including classical BN and three recently published quantum-like decision models [28][29][42], as well as the proposed PEQBN.

The proposed PEQBN model draws upon the strength of the basic ideas of its predecessors while avoiding their weaknesses and drawbacks in modeling human selection behavior. Specifically, we consider each DM as part of a whole multi-agent system instead of an independent agent. So, all relationships due to unknown sources in society may affect human decisions. Our innovation here is introducing an *entangled* quantum-like BN structure in which nodes are simulated as photons related to each other by some unknown links in addition to causal relations. In the presence of uncertainty, photons behave as entangled waves instead of isolated particles. Hence, we consider an initial superposition state of choices for each DM formed by the environmental condition in the society. The dynamical evolutions of selecting all possible choices in the proposed structure are modeled by entangled wave functions that demonstrate interference effects due to overlapping different decision-making processes.

In quantum systems, entanglement measures and witnesses are used to measuring the stochastic unknown long-range effect. We apply a similar tool to model the impact of society on human selection behavior. We estimate the interference term by applying two famous entanglement measures, including Shannon entropy and concurrence, as well as a proposed quantum-like witness ($qW$) in this paper. So, the estimation of the interference effects in this study is based on the physical concepts of quantum information theory rather than heuristic functions, and the entanglement stands as the central idea.

The proposed model is evaluated on 22 experiments in three categories: four experiments of prisoner's dilemma, three experiments of two-stage gambling game, and 15 experiments of document relevance judgment. In all tasks, predicted probabilities are obtained by our model and four predictive decision-making models, including classical BN, quantum decision theory, and two versions of predictive quantum-like Bayesian networks. The deviations between predicted probabilities and empirical results of human decisions reveal that the proposed model is superior in predicting human decisions under uncertainty.

In the future, we hope to use more data from human agents for a more realistic evaluation and model identification. We also plan to extend our model to simulate the decision-making tasks with more than two choices by entangled BNs. Then we can apply PEQBN in more complex human decision-making challenges in various fields such as crisis management, traffic control, or information diffusion in social networks.

APPENDIX 1. THE DEFINITION AND CALCULATION METHOD OF CONCURRENCE MEASURE [58]

$$C(\rho_x) = max\{0, \lambda_1 - \lambda_2 - \lambda_3 - \lambda_4\}, \quad \lambda_1 \geq \lambda_2 \geq \lambda_3 \geq \lambda_4$$
$$\lambda_i \text{ is the eigenvalue of } Matrix_C$$
$$Matrix_C = \rho_x * Y \otimes Y * \rho_x^\dagger * Y \otimes Y$$
$$Y = \begin{bmatrix} 0 & -i \\ i & 0 \end{bmatrix} \text{ is the second Pauli matrix}$$

| $\rho_x = \|\psi_x><\psi_x\|$ | $\|\psi_x> = (c_1 e^{i\theta_1} \quad c_2 e^{i\theta_2} \quad c_3 e^{i\theta_3} \quad c_4 e^{i\theta_4})'$ |
|---|---|
| | In 4 dimension Hilbert space |